\documentclass[aps,pra,preprint]{revtex4-1}
\pdfoutput=1
\usepackage[latin1]{inputenc}
\usepackage[english]{babel}
\usepackage{latexsym}
\usepackage{graphicx}
\usepackage{amsmath}
\usepackage{natbib}

\newcommand{\rp}{$\langle \mathrm{r_p}\rangle$}

\newcommand{\Hep}{He-e$^+$}

\newcommand{\Veff}{$\mathrm{V_{eff}^{Ps}}$}
\newcommand{\Veffnorm}{$\mathrm{V_{eff}^{'Ps}}$}

\newcommand{\singlet}{$\mathrm{^1S_e}$}
\newcommand{\triplet}{$\mathrm{^3S_e}$}
\newcommand{\po}{$\mathrm{\lambda^{po}_c}$}
\newcommand{\latticeparam}{$\mathrm{a_{lat}}$}
\newcommand{\PhaseShift}{$\delta_0$}
\newcommand{\ScatLength}{$A_0$}
\newcommand{\rsph}{$r_{sph}$}

\begin{document}

\title{
Pick-off annihilation of positronium in matter using 
full correlation single particle potentials: solid He
}
\author{A. Zubiaga}
\email{asier.zubiaga@gmail.com}
\author{F. Tuomisto}
\affiliation{Department of Applied Physics, Aalto University, P.O. Box 14100, FIN-00076 Aalto Espoo, Finland}
\author{M.~J. Puska}
\affiliation{COMP, Department of Applied Physics, Aalto University, P.O. Box 11100, FIN-00076 Aalto Espoo, Finland}

\begin{abstract}
We investigate the modeling of positronium (Ps) states and their pick-off annihilation trapped at open volumes pockets in condensed molecular matter. Our starting point is the interacting many-body system of Ps and a He atom because it is the smallest entity that can mimic the energy gap between the highest occupied and lowest unoccupied molecular orbitals of molecules and yet the the many-body structure of the HePs system can be calculated accurately enough. The exact-diagonalization solution of the HePs system enables us to construct a pair-wise full-correlation single-particle potential for the Ps-He interaction and the total potential in solids is obtained as a superposition of the pair-wise potentials. We study in detail Ps states and their pick-off annihilation rates in voids inside solid He and analyse experimental results for Ps-induced voids in liquid He obtaining the radii of the voids. More importantly, we generalize our conclusions by testing the validity of the Tao-Eldrup model, widely used to analyse ortho-Ps annihilation measurements for voids in molecular matter, against our theoretical results for the solid He. Moreover, we discuss the influence of the partial charges of polar molecules and the strength of the van der Waals interaction on the pick-off annihilation rate.
\end{abstract}


\maketitle

\section{Introduction}
An appreciable fraction of positrons implanted inside molecular soft-condensed matter (polymers, liquids) form positronium (Ps). The Ps atom is the bound state of a positron and an electron and its relevance comes from its distinctive chemical properties. Thermalized Ps gets localized at open volume pockets such as vacancies or voids where the Ps-matter repulsion is minimum. Ortho-Ps (o-Ps) is the long-lived (142~ns) spin-triplet state of Ps but inside matter it annihilates mainly through a faster pick-off process with an electron of the host material. During the annihilation process two gamma photons are emitted and o-Ps lifetime is reduced to 1-50~ns depending on the local electronic structure and the size of the open volume pocket~\cite{Book_Mogensen}. In interaction with radicals or atoms with unpaired electrons, o-Ps can form a strong chemical bond (chemical quenching) or become a para-Ps spin-singlet (spin conversion). In both cases the lifetime of o-Ps is dramatically reduced below 1~ns. The experimental signature of the pick-off annihilation of o-Ps can, thus, be separated from other annihilation processes where the positron interacts stronger with the electrons of matter. Positronium annihilation lifetime spectroscopy exploits this property to measure the distribution of the open volume in materials such as porous SiO$\mathrm{_2}$~\cite{PRB_Nagai, APL_Liszkay}, polymers~\cite{JPS_Uedono} or biostructures~\cite{JPCBL_Sane, JPCBL_Dong}. 

The analysis of experiments benefits from predictions by computational models. However, an ab-initio treatment of a system including two correlated light-particle species with an attractive interaction such as the electron and the positron is a heavy computational problem. In metals and semiconductors, where Ps does not form, the two-component density functional theory gives accurate predictions for the positron annihilation parameters~\cite{RMP_Puska, RMP_Tuomisto}. But for soft molecular matter there is no practical ab-initio scheme for an efficient predictive description of the properties of Ps. 

In this work we study the interacting He-Ps system whose structure can be solved accurately with quantum many-body ab-initio techniques. Moreover, the similarities of the He electronic structure with that of the molecular matter, i.e., it possesses an energy gap between the highest occupied molecular orbital (HOMO) and the lowest unoccupied molecular orbital (LUMO) in a spin-compensated electron structure, makes it a good model system to study in detail the Ps-material interaction. The quantum many-body ab-initio results including the exact correlations of the interacting system can be used to derive a full-correlation effective potential for the interaction between He and Ps. On the other hand, we will have to be careful considering the effects of the low polarizability of He when extrapolating our results to typical soft-matter.

In practice, we calculate the many-body wavefunction of HePs using the exact diagonalization stochastic variational method (SVM) and an explicitly-correlated Gaussian (ECG) function basis set. This method has given the lowest total energies for few particle systems including a positron~\cite{PRC_Varga,PRA_Bubin,JAMS_Mitroy}. Thereafter, from the many-body positron density we derive a single particle positron potential for Ps interacting with He similarly to Ref.~\cite{PRA_Zubiaga2}. Then we use the superposition of the ensuing atomic potentials to calculate the total potential in solid He and solve the single-particle Schr\"odinger equation. Thereby, we can calculate the distribution of the positron in Ps which allows us to discuss the distribution of Ps and its pick-off annihilation rate in the bulk and in voids of different sizes and geometries. The pick-off annihilation rates are in qualitative agreement with the semi-empirical Tao-Eldrup (TE) model~\cite{JCP_Tao, CP_Eldrup}. Our approach takes into account the geometry and size effects as well as the chemical specificity of the material. We also address the influence of the van der Waals interaction and the partial charges of polar molecules in the pick-off annihilation properties of Ps. 

The organization of the present paper is as follows. The many-body ECG-SVM, the construction of the single-particle atom-Ps potential, as well as the practical scheme to calculate Ps states and annihilation rates in condensed matter are shortly described in Chapter~\ref{sec2}. Moreover, Chapter~\ref{sec3} discusses the behaviour of the He-Ps potential in detail and how it is affected by modifying the nuclear charge to mimic polar molecules. Chapter~\ref{sec_spherical} contains the main results of our work, i.e., the calculated Ps distributions and pick-off annihilation rates at voids in solid He as well as the discussion on the validity of the TE model. Finally, chapter~\ref{sec5} is a short summary.

\section{Methods}\label{sec2}
\subsection{ECG-SVM method} 
We use the ECG-SVM~\cite{PRC_Varga} all-particle quantum ab-initio approach, to calculate many-body wavefunctions and total energies of systems composed by a nucleus, N electrons, and a positron interacting through the Coulomb force. The hadronic nucleus is considered as structureless, on equal footing with the electrons and the positron. The wavefunction is expanded in terms of a linear combination of properly antisymmetrized s-type ECG functions, since all the systems that we have considered have zero orbital angular momentum, i.e., 
\begin{eqnarray}\label{eq1}
\Psi(x) 
= \sum_{i=1}^s c_i\ 
{\displaystyle \mathcal{A} } 
\left [ 
\exp
^{-\frac{1}{2}xA^ix}
\right ]
\otimes\chi_{SMs} 
.
\end{eqnarray}
$A^i$ are the non-linear coefficient matrices, $c_i$ are the mixing coefficients of the eigenvectors and $\chi_{SMs}$ is a spin eigenfunction. The antisymmetrization operator $\mathcal{A}$ acts on indistinguishable particles and the Jacobi coordinate sets \{$x_1$,...,$x_{N-1}$\} with reduced masses $\mu_i$ = $m_{i+1}\sum_{j=1}^im_j/\sum_{j=1}^{i+1} m_j$ allow for a straightforward separation of the centre of mass (CM) movement. The electron and positron densities are determined as $n_-(r) = \sum_{i=1}^{N_e}\langle\Psi |\delta(\vec{r}_i-\vec{r}_N-\vec{r})|\Psi\rangle$ and $n_{+}(r) = \langle\Psi |\delta(\vec{r}_p-\vec{r}_N-\vec{r})|\Psi\rangle$, respectively. Here $\vec{r}_i$, $\vec{r}_p$, and $\vec{r}_N$ are the coordinates of the i$^{th}$ electron, the positron and the nucleus, respectively.  

The ECG basis sets comprise between 600 and 2000 functions according to the size of the system. The non-linear coefficients $A^i$ need to be optimized to avoid very large basis sets. In the SVM large parameter vectors are optimized sequentially by giving them random values which are kept only if the update lowers the total energy of the system. This is a time consuming procedure but the success of the ECG-SVM method relies on the efficient calculation of the matrix elements. 

The wavefunctions are eigenstates of the non-relativistic Hamiltonian with the kinetic energy of the CM subtracted, i. e.,  
\begin{equation}
H = \sum_i\frac{p_i^2}{2m_i} - T_{CM} + \sum_{i<j}\frac{q_iq_j}{r_{ij}},
\end{equation}
where $\vec{p}_i$ is the momentum, $m_i$ the mass, and $q_i$  the charge of the i$^{th}$ particle, $r_{ij}$ is the distance between the $i^{\rm th}$ and $j^{\rm th}$ particles and $T_{CM}$ the kinetic energy of the CM. 
\begin{table*}[h!]
\caption{
Main properties of the calculated systems. The components, the size of the basis, the total energy, the mean positron-nucleus distance \rp{}, and the Ps interaction energy (E$^{Ps}_{int}$) are given. The average confinement potential energy $\mathrm{\langle V_{conf}\rangle}$ has been subtracted from the interaction energies for \triplet{}-HPs and HePs.}
\label{tab1}
\begin{ruledtabular}
\begin{center}
\begin{tabular}{lcccc}
&Basis size&Energy&\rp{}&E$^{Ps}_{int}$\\
&(\# functions)&(a.u.)&(a.u.)&(a.u.)\\
\hline
He&600&-2.9037\\
\singlet{}-HPs&1000&-0.78919&3.66&-0.39187$\times$10$^{-1}$\\
\triplet{}-HPs&2000&-0.74991&88.74&0.8662$\times$10$^{-4}$\\
HePs&2000&-3.15351&59.37&0.1191$\times$10$^{-3}$\\
\end{tabular}
\end{center}
\end{ruledtabular}
\end{table*}

In order to calculate unbound \triplet{}-HPs and HePs we add a two-body confining potential binding the positron to the hadronic nucleus similar to the potential used by Mitroy et al.~\cite{PRL_Mitroy}, 
\begin{equation}
V_{conf}(r_p) = \left\{
\begin{array}{l@{\ \ ,\ \ \ }r}
0&r<R_0\\
\alpha(r_p-R_0)^2&r\ge R_0.
\end{array}
\right.
\end{equation}
The potential is only non-zero beyond the confinement radius $R_0$. Coefficients $R_0$ and $\alpha$ are chosen so that \rp{} $>$ 50~a.u. to prevent the wavefunction to be affected by the confinement potential within the Ps-atom interaction region. Resulting values for the average confining potential $\mathrm{\langle V_{conf}\rangle}$ = 1-3$\times$10$^{-6}$~a.u. are lower than E$^{Ps}_{int}$ in all cases. 

We define the Ps interaction energy, $E^{Ps}_{int} = E_{XPs} - E_{X^+} - E_{Ps}$, as the difference between the total energy of the interacting system $E_{XPs}$ and the sum of the total energies of the neutral atom $E_{X}$ and Ps $E_{Ps}$. For the confined states the confinement potential energy $\langle V_{conf}\rangle$ has been subtracted from $E^{Ps}_{int}$. Detailed information on the calculated systems is given in Table~\ref{tab1}. 

\subsection{Construction of single-particle atom-Ps potentials and modeling of Ps in condensed matter}
We define \Veff{}$\textrm{(r)}$, where the distance $r$ is measured from the nucleus, as in Ref.~[\onlinecite{PRA_Zubiaga2}], i.e., by inverting the single-particle Schr\"odinger equation with the square root of the positron density $n^+(r)$ as the eigenfunction and the interaction energy as the energy eigenvalue,  
\begin{equation}\label{eq_def_veff}
V_{eff}^{Ps}(r) = E_{int}^{Ps} + \frac{1}{2M^{Ps}}\frac{\nabla^2\sqrt{n^+(r)}}{\sqrt{n^+(r)}}. 
\end{equation}
$E_{int}^{Ps}$ is given in table~\ref{tab1} and $M^{Ps}$=2$m_e$ is the mass of Ps. \Veff{} is a single-particle positron potential for systems forming a Ps-like subsystem. The positron $\mathrm{V_{eff}}$ is defined similarly to Eq.~(\ref{eq_def_veff}) for systems were a Ps-like subsystem doesn't form~\cite{PRA_Zubiaga2}. The introduced potential is equivalent to the exact Kohn-Sham potential for a single positron. 

The asymptotic value of \Veff{} is zero far from the nucleus by construction. The single particle Schr\"odinger equation solved with \Veff{} and $M^{Ps}$=2$m_e$ recovers the Ps interaction energy in confined states and the Ps binding energy in bound states~\cite{PRA_Zubiaga2}. Alternatively, for positrons forming Ps the potential $\mathrm{V_{eff}^{'Ps} = 2 E_{int}^{Ps} + \nabla^2(\sqrt{n_+})/(2\sqrt{n_+})}$ with effective mass $m_e$ can be defined. It yields the same positron density as \Veff{} but the energy eigenvalue is multiplied by a factor of 2. In this work, \Veffnorm{} will be useful in comparisons to the mean-field Coulomb potentials and the positron $\mathrm{V_{eff}}$. 

We use the superposition of the single particle Ps-He interaction potentials \Veff{} to calculate Ps states trapped inside voids in solid $^4$He. The total potential of Ps inside a solid is calculated as the superposition of \Veff{} for a single atom. This is a good approximation when the electronic structures of isolated atoms, or in a general case of isolated molecules, are not substantially modified in the condensed state. The density and the ground state energy inside a void are calculated by solving the single particle Schr\"odinger equation for the superposition potential with an effective mass $2m_e$. 

In practice the problem is discretized on a three-dimensional real-space grid and solved by using a numerical relaxation technique and periodic boundary conditions~\cite{JPF_Puska}. For large voids we take special care that the positron density is small at the supercell boundary so that the interactions between periodic images are negligible. 

The pick-off annihilation rate (\po{}) of Ps can be calculated from the overlap integral of the positron and electron densities~\cite{PRA_Zubiaga} as
\begin{equation}\label{eq_lambda}
\mathrm{\lambda^{po}_c} = \pi r_0^2c\int n_+(\vec{r}) n_-(\vec{r}) d\vec{r}, 
\end{equation}
where $r_0$ is the classical electron radius, $c$ is the speed of light, $n_+(r)$ is the positron density, and $n_-(r)$ is the total electron density obtained by superimposing atomic ECG-SVM electron densities. 

\section{Single-particle interaction potentials}\label{sec3}
\begin{figure*}
\begin{center}
\includegraphics[width=16cm]{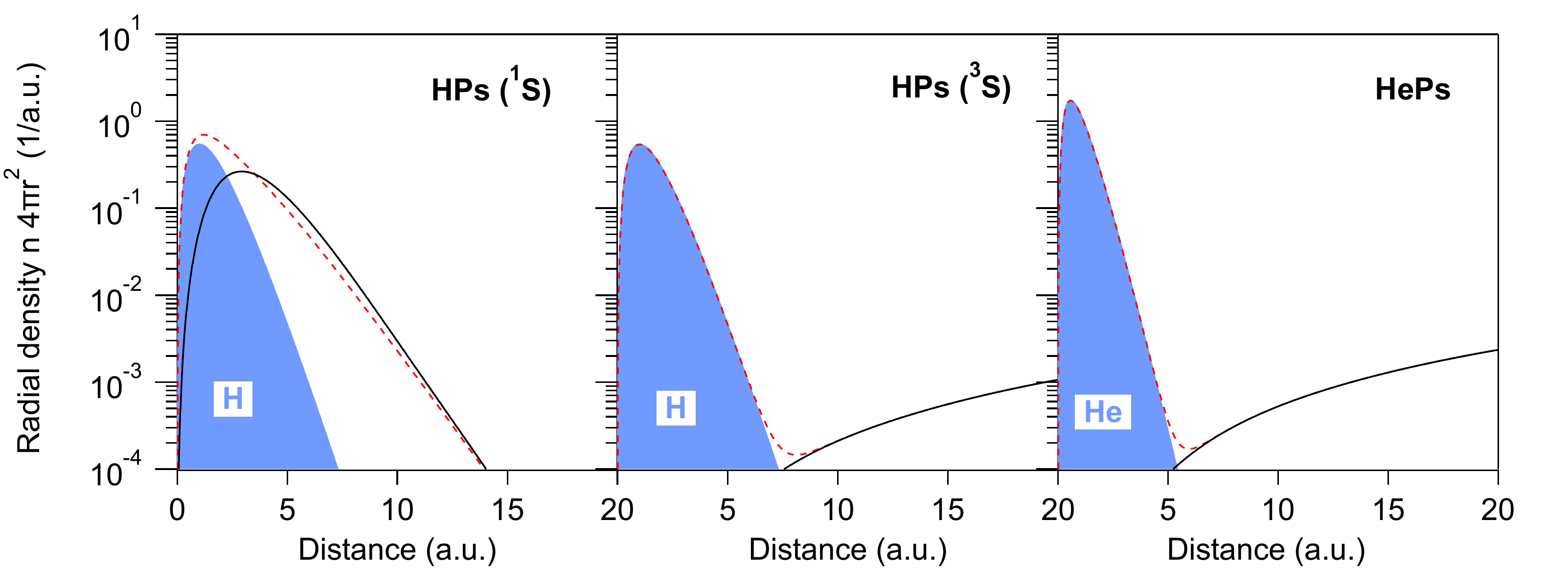}
\caption{(Color online) Electron and positron densities of \singlet{}-HPs, \triplet{}-HPs, and HePs. The electron densities of the isolated atoms (filled blue curves), and the interacting positron-atom systems (red broken curves), as well as the positron densities (black full curves) are shown.} 
\label{fig1}
\end{center}
\end{figure*}

The ECG-SVM particle densities of \singlet{}-HPs, \triplet{}-HPs and HePs are plotted in Fig.~\ref{fig1}. In \singlet{}-HPs the electron of Ps forms a spin singlet with the electron of H and the positron enters the atom electron cloud despite of the repulsion exerted by the nucleus. On the other hand, in the weakly interacting \triplet{}-HPs and HePs the electron density of the atoms remain largely undisturbed and the large repulsion felt by the electron and the positron of Ps prevent them from entering the electron cloud of the atom. They instead remain bound in an unpolarized Ps state. 

\begin{figure*}
\begin{center}
\includegraphics[width=16cm]{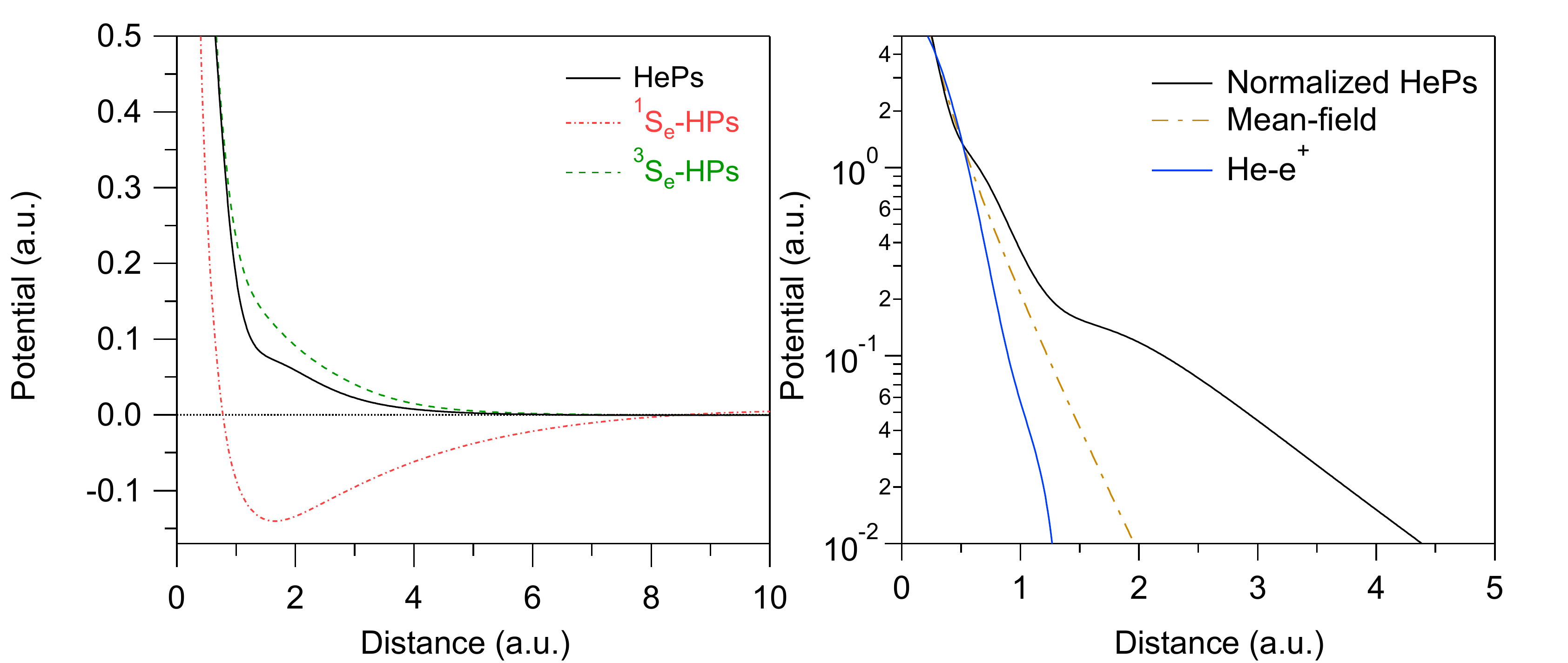}
\caption{(Color online) \Veff{} for the calculated systems. The left panel shows \Veff{} for HePs (black full curve), \singlet{}-HPs (light red dash-dotted curve), and \triplet{}-HPs (light green dashed curve). The right panel shows \Veffnorm{} of HePs, the positron mean-field Coulomb potential of He (brown dash-dotted line) and the $\mathrm{V_{eff}}$ of \Hep{} (blue line).} 
\label{fig2}
\end{center}
\end{figure*}

He doesn't bind Ps~\cite{PRA_Mitroy6} due to its closed shell structure and low polarizability. HePs is the smallest system where Ps interacts with a molecule or an atom having a HOMO-LUMO gap. \Veff{} for HePs, shown in the left panel of Fig.~\ref{fig2}, remains purely repulsive until 8~a.u. The positron \Veff{} of the unbound electronic triplet state \triplet{}-HPs also decays slowly to zero but for the bound electronic singlet state \singlet{}-HPs \Veff{} has a deep binding potential at 1.5~a.u. The long-range repulsive tails of \triplet{}-HPs and HePs reflect the electron-electron repulsion felt by the electron of Ps and the confinement kinetic energy of the light Ps atom~\cite{JPCS_Zubiaga}. 

The right panel of Fig.~\ref{fig2} illustrates how \Veffnorm{} of HePs is dominated by the mean-field Coulomb positron potential (without the electron-positron correlation) close to the nucleus, r~$<$~1~a.u., where the positron-nucleus Coulomb repulsion dominates. In \Hep{} the positron-nucleus Coulomb repulsion is also dominant close to the nucleus. However, the exchange repulsion doesn't play any role and $\mathrm{V_{eff}}$, plotted in the right panel of Fig.~\ref{fig2}, lacks a long-range repulsive tail. Instead it has a shallow attractive well for separations from the nucleus of 1.3~a.u.~\cite{PRA_Zubiaga2}. 

We consider now low energy Ps scattering off He in order to study the adequacy of \Veff{} to model Ps states. At low energies the scattering properties are well described by the s-wave phase shifts (\PhaseShift{}) and scattering lengths (\ScatLength{}). Zhang et al.~\cite{PRA_Zhang2} used stabilized ECG-SVM to calculate \PhaseShift{} and \ScatLength{}. We have calculated \PhaseShift{} and \ScatLength{} using the single-particle potential \Veff{}. The s-wave scattering wavefunction for the positron in Ps is the solution to the radial single-particle Schr\"odinger equation 
\begin{equation}\label{eq2}
-\frac{1}{2M_{eff}}\frac{d^2 U}{dr^2} + V_{eff}U = EU,
\end{equation}
where $U = r\psi_0$, $\psi_0$ is the s-type wavefunction and $E=k^2/2M_{eff}$ the energy of Ps. $U$ obeys the boundary conditions $U(r=0)$=0 at the origin and at large distances from the nucleus the solution has the form 
\begin{equation}\label{eq_scatt_wavefunc}
\lim_{r\longrightarrow\infty} \psi_0 = \frac{\sin\left(kr+\delta_0\right)}{kr}. 
\end{equation}
We calculate \PhaseShift{}($k$) by fitting the wavefunction calculated numerically to this asymptote and \ScatLength{} is calculated at the low-energy limit using $k \cot\delta_0 = -1/A_0 + O(k^2)$. The \Veff{} value of \ScatLength{} is 1.3~a.u., in a fairly good agreement with the many-body value $A_0^{ECG}$ = 1.566~a.u. 

\begin{figure*}
\begin{center}
\includegraphics[width=8.5cm]{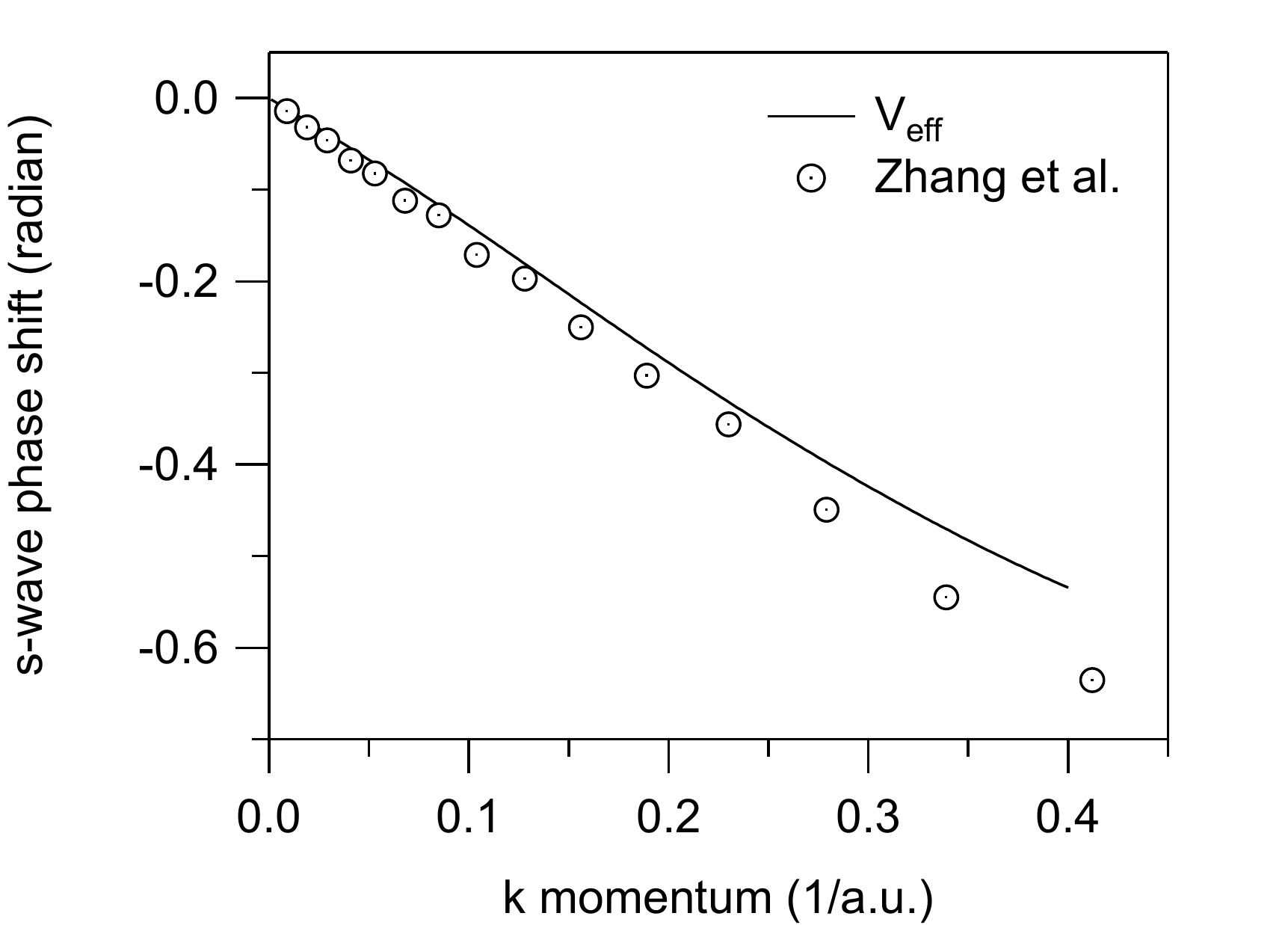}
\caption{Phase shifts of the scattering s-wave. The values calculated using \Veff{} (black line) are compared to the many-body values from Zhang et al.~\cite{PRA_Zhang2} (dotted circles). } 
\label{fig7}
\end{center}
\end{figure*}

The corresponding \PhaseShift{} as a function of $k$, shown in Fig.~\ref{fig7}, is slightly larger than the many-body results. The agreement is very good for E $\leq$ 2.5$\times$10$^{-3}$~a.u. (k $\leq$ 0.1~1/a.u.), which is clearly beyond the thermal energy of Ps at room temperature (E = 0.95$\times$10$^{-3}$~a.u.). 

\subsection{Single-particle potentials for \uppercase{H}\lowercase{e}-like ions}
We have considered He-like ions with nuclear charges ranging between Z=1.5 and Z=2.5. Our motivation is to discuss the effect that partial charges of polar molecules have on the Ps distribution and pick-off annihilation rate. All the ion-Ps systems remain unbound and the mean nucleus-positron distances and the Ps interaction energies, without the confinement energy, remain fairly constant, as shown in table~\ref{tab2}. On the other hand, the pick-off annihilation rate is 1.2$\times$10$^{-3}$~ns$^{-1}$ for Z=1.5 and decreases to 7.5$\times$10$^{-5}$~ns$^{-1}$ for Z=2.5, the most positive ion. 
\begin{table*}[h]
\caption{Total energies of the interacting HePs-like systems $_{\rm Z}$He-Ps, the isolated ions $_{\rm Z}$He and the corresponding Ps interaction energy $E_{int}^{Ps}$ for nuclear charges Z=1.5, 2.0 and Z=2.5. Atomic units are used for all the magnitudes except for the pick-off annihilation rate that is given in ns$^{-1}$.}
\label{tab2}
\begin{ruledtabular}
\begin{center}
\begin{tabular}{cccccc}
Z&$_{\rm Z}$He-Ps&$_{\rm Z}$He&$E_{int}^{Ps}$&\rp{}&$\Gamma^{po}$\\
\hline
1.5&-1.71484&-1.46526&0.0004262&34.87&12.17$\times$10$^{-4}$\\
2&-3.15321&-2.90369&0.000481625&36.37&1.14$\times$10$^{-4}$\\
2.5&-5.09136&-4.84187&0.00050481&33.15&0.752$\times$10$^{-4}$\\
\end{tabular}
\end{center}
\end{ruledtabular}
\end{table*}
\begin{figure*}
\begin{center}
\includegraphics[width=16cm]{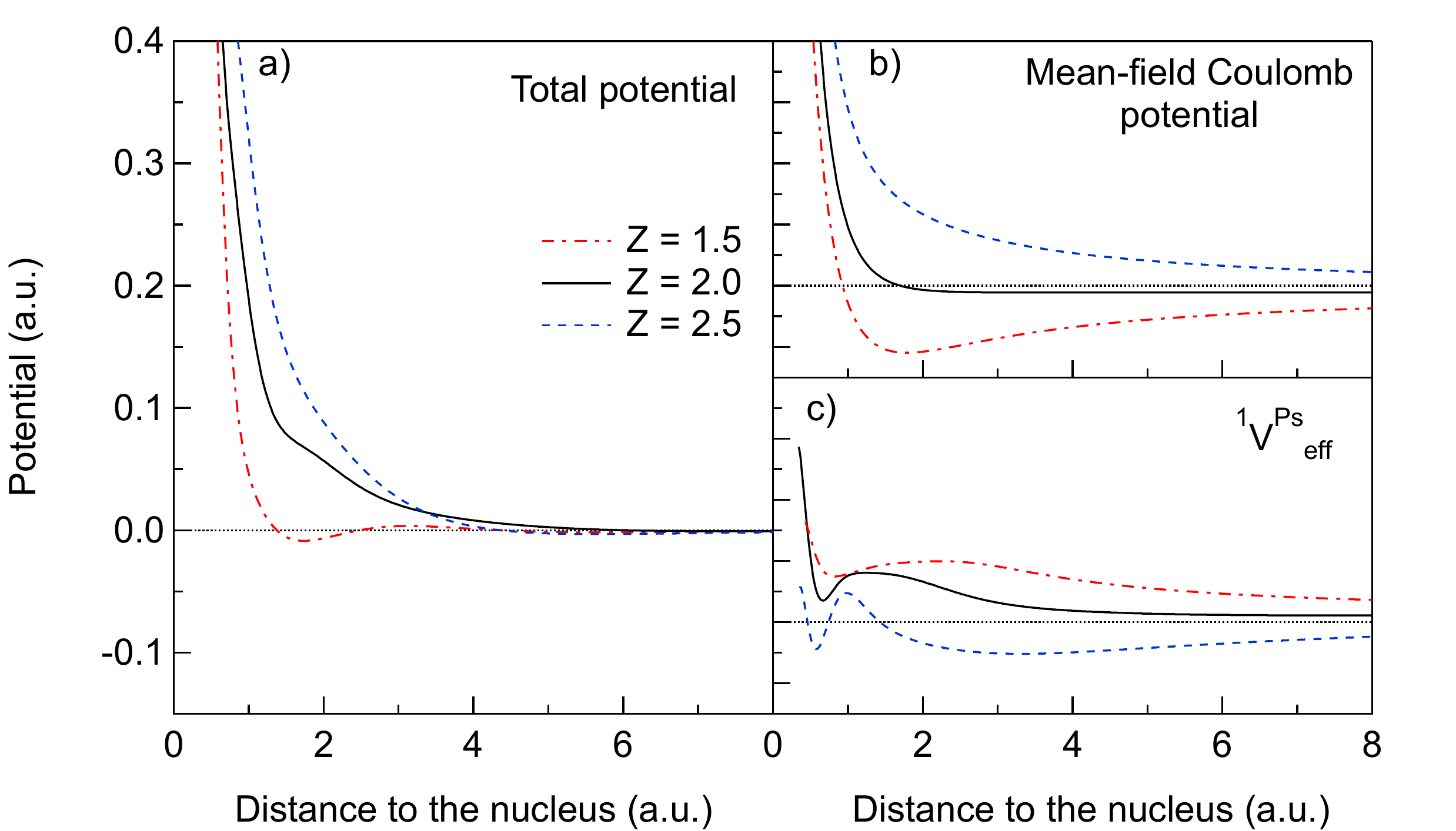}
\caption{(Color online) Full correlation potentials \Veff{} of He-like ions with Z=1.5 (red dashed-dotted curve), Z=2 (black curve), and Z=2.5 (blue dashed curve): a) Total potentials, b) mean-field Coulomb potentials, and c) $\mathrm{^1V_{eff}^{Ps}}$.} 
\label{fig11}
\end{center}
\end{figure*}

The corresponding \Veff{} are shown in the left panel of figure~\ref{fig11}. Overall, the ions with larger Z repel stronger the positron. The mean-field Coulomb repulsion is the dominant contribution close to the nucleus and it determines the size of the repulsive core. The many-body effects are described by $\mathrm{^1V_{eff}^{Ps}}$ = \Veff{} - $\mathrm{V_C}$ subtracting the mean-field Coulomb potential $\mathrm{V_C(r) = Z/r-\int d\vec{r}' n_-(r')/|\vec{r}-\vec{r}'|}$. $\mathrm{^1V_{eff}^{Ps}}$, shown in the right panel of Fig.~\ref{fig11}, includes the electron-electron exchange and correlations and the electron-positron correlations. The net effect of the electron-electron and electron-positron correlations is a dispersion attraction between the ion and Ps which depends on the electronic properties of the ion. In ions with low Z the electron-electron exchange repulsion dominates over the fast decaying dispersive interaction and $\mathrm{^1V_{eff}^{Ps}}$ is repulsive. Contrary, thanks to the compact electron cloud of the ions with high Z the dispersion interaction dominates over the electron-electron repulsion. 

The variations in the spatial extents of the electron clouds of different Z forces us to be cautious when extrapolating our results to polar molecules. However, our results suggest that the dispersion interaction can play an important role for polar molecules. Ps will pile-up close to the negative partial charges with the positron facing the molecule. A similar result has already been observed for a positron interacting with alkali-metal hydrides~\cite{JCP_Kita2}. Importantly, the resulting pick-off annihilation rate is expected to be enhanced compared to homopolar molecules. 
\begin{figure*}
\begin{center}
\includegraphics[width=16cm]{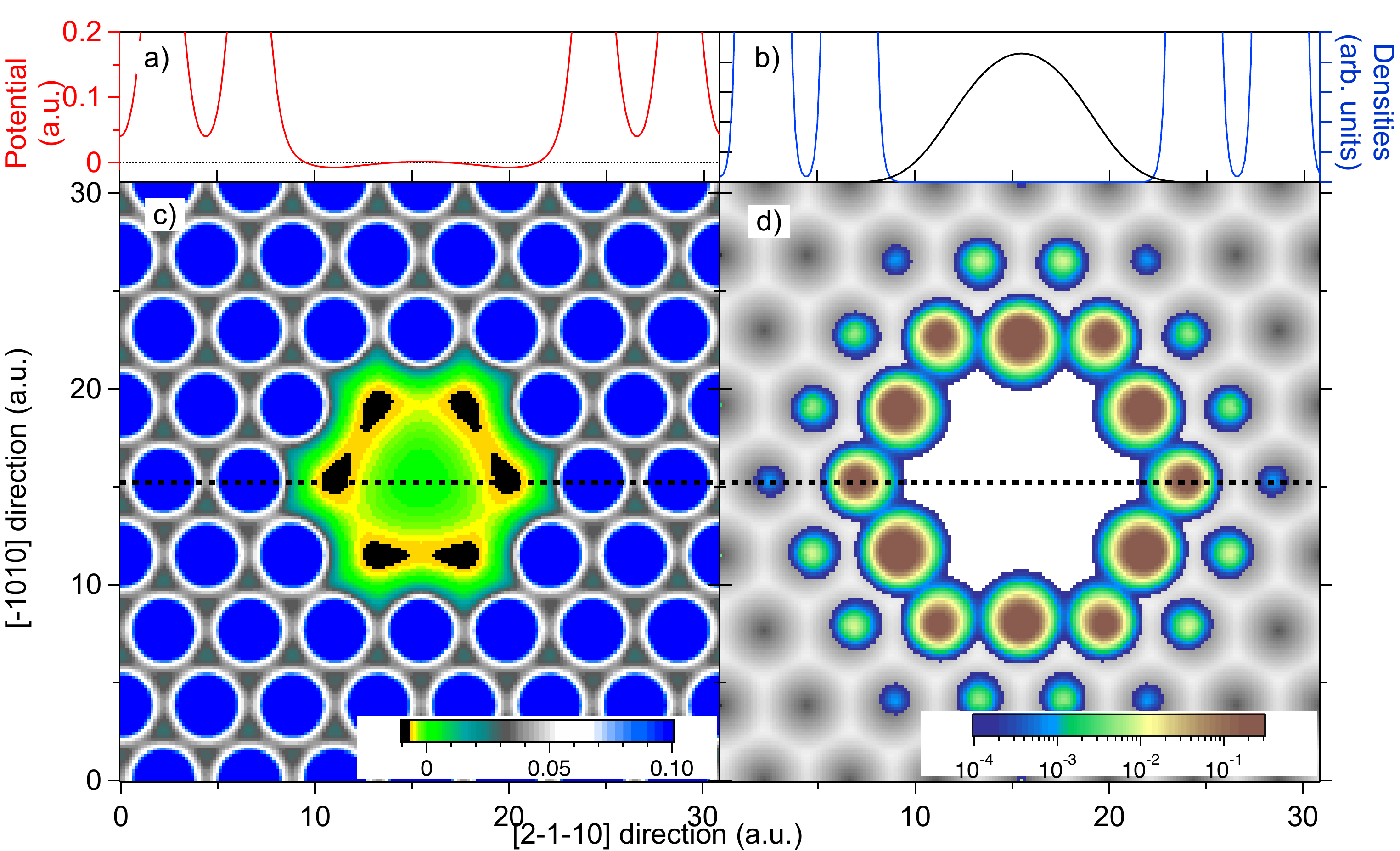}
\caption{(Color online) Ps inside a spherical void in solid He formed by 19 missing atoms (\latticeparam{}=8.3~a.u., \rsph=12.3~a.u.). The upper panels show a) the positron potential (red curve), b) the positron (black curve) and electron (blue curve) densities plotted along the [2-1-10] direction through the centre of the void. Panel c) shows the positron potential (a.u.) and panel d) the electron density (gray shade scale) and the local annihilation rate (1/ns) in the (0001) plane.} 
\label{fig10}
\end{center}
\end{figure*}

\section{Voids in solid H\lowercase{e}}\label{sec_spherical} 
$^4$He crystallizes in a hexagonal close packed solid phase below 15~K when compressed above 25 bar~\cite{Book_Glyde}. Due to its large compressibility the lattice parameter vary between 8.3~au at 35~bar and 4.0~au at 4.9~kbar. Ryts\"ol\"a et al.~\cite{JPB_Rytsola} measured the lifetime of o-Ps in solid He. They found that Ps annihilates in He inside a void ("bubble") originated by the He-Ps short range repulsion. The measured pick-off annihilation rates range between 0.012~1/ns (25~bar) and 0.018~1/ns (63~bar). 
\begin{figure}
\begin{center}
\includegraphics[width=8.5cm]{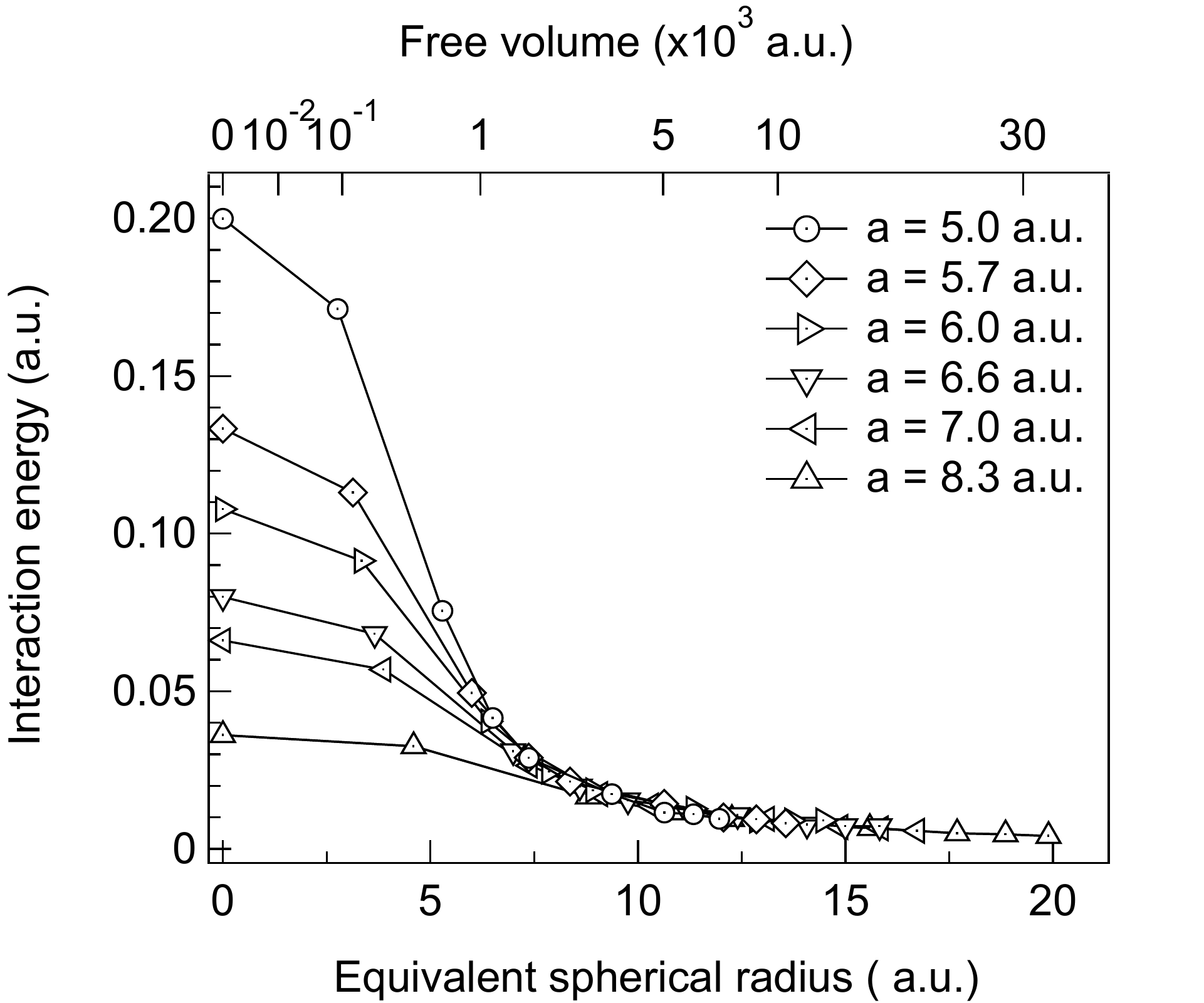}
\caption{(Color online) Total Ps-matter interaction energy in bulk solid He (leftmost markers) and in spherical voids for different lattice parameters plotted versus the equivalent spherical radius of the void. The top axis shows the corresponding free volume of the void. 
} 
\label{fig9}
\end{center}
\end{figure}

We have calculated the positron distribution for Ps in six rectangular supercells comprising 448 He atoms each with lattice parameters (\latticeparam{}) between 5.0~a.u and 8.3~a.u. We have introduced voids of different sizes and shapes with open volumes ranging from 1 to 126 missing He atoms. The free volume of the void ($V_{free}$) was estimated as the the sum of the specific volumes ($V_{sc}$/$N_{sc}$) of the missing atoms ($N_{vac}$): $V_{free} = V_{sc}N_{vac}/N_{sc}$. $V_{sc}$ and $N_{sc}$ are the total volume and number of atoms of the supercell, respectively. For spherical voids, we use the equivalent spherical radius \rsph{} = $\left(3V_{free}/4\pi\right)^{1/3}$ instead of $V_{free}$. 

\subsection{Spherical voids}
Tao~\cite{JCP_Tao} proposed a phenomenological model of Ps trapped in spherical voids widely used to predict the sizes of the voids from measured pick-off annihilation rates. The trapping potential is approximated as an infinite spherical potential well and Ps is treated as a single particle. The wavefunction can be calculated analytically and the pick-off annihilation rate ($\lambda^{po}_{TE}$) of Ps in a spherical void of radius R is calculated as the overlap with a electron density layer of constant density and thickness $\Delta R$ in the wall of the void, 
\begin{equation}\label{eq_gammaTE}
\lambda^{po}_{TE} = 2.0 (\textrm{ns}^{-1})\left[ 1-\frac{R}{R+\Delta R}+\frac{1}{2\pi}\sin\left(\frac{2\pi R}{R+\Delta R}\right) \right].
\end{equation}
The values of $\Delta$R have been estimated to lie between 1.66~\AA (3.14~a.u.) and 1.9~\AA (3.59~a.u.) in plastic crystals, liquids and zeolites~\cite{CPL_Goworek, CP_Eldrup}. The model has been extended to include rectangular voids~\cite{JRNC_Jasinska} and excited states in large voids~\cite{CP_Goworek}. 

We have studied Ps trapped in (quasi-)spherical voids with sizes ranging between 1 (monovacancy) and 81 missing atoms. 
The total potential in panels a) and c) of figure~\ref{fig10} is minimum inside the void. The dispersion attraction induces a shallow potential well near the wall of the void. The pick-off annihilation rate in panel d) results from the the overlap of the positron and electron densities, as illustrated in panel b). The positron pick-off annihilates mostly with the atoms forming the wall of the void although also the second neighbours contribute. The probability to annihilate inside the void is very low because the electron density, gray shade in panel d), is negligible there. 
\begin{figure}
\begin{center}
\includegraphics[width=8.5cm]{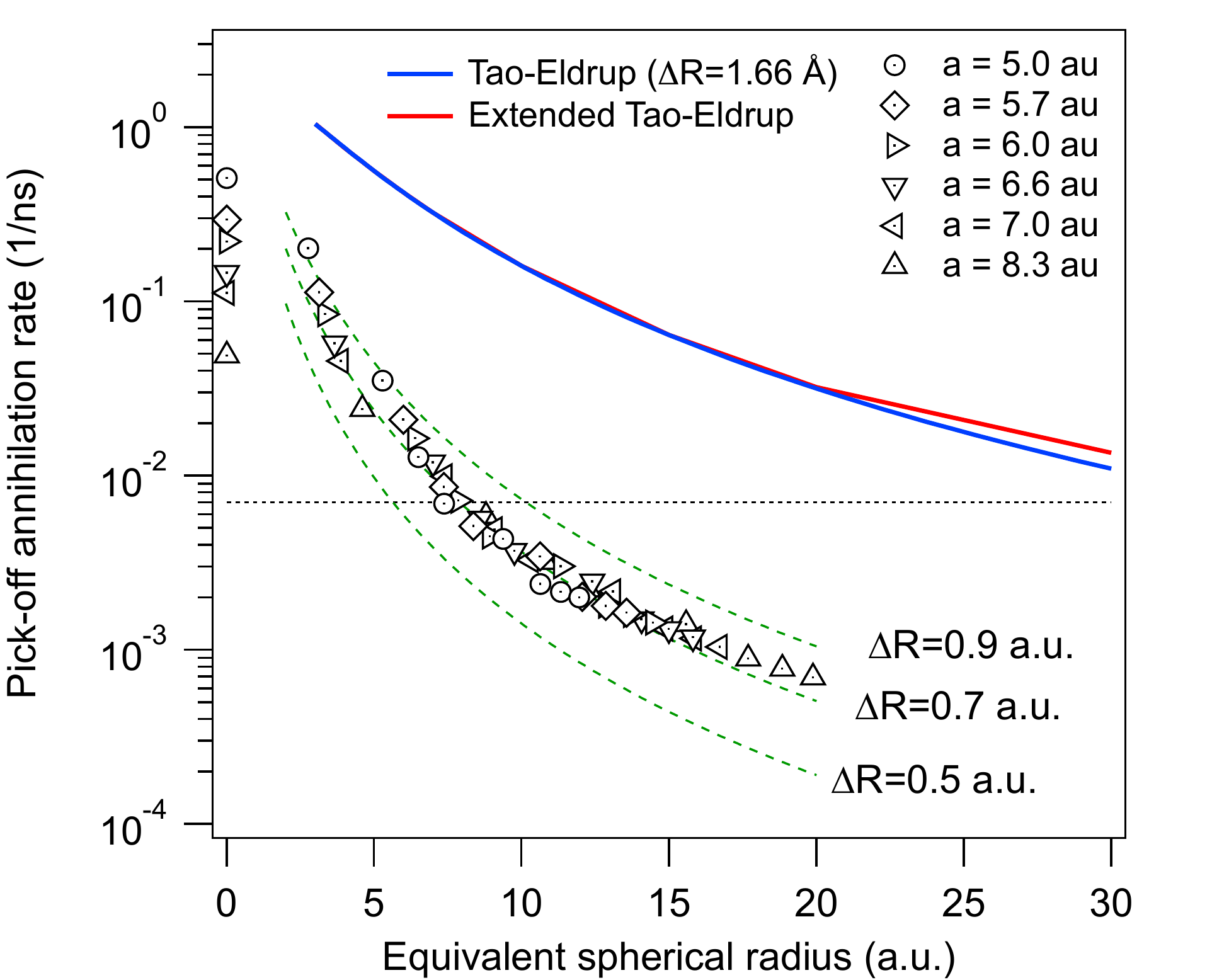}
\caption{(Color online) Ps pick-off annihilation rates for voids in solid He of different densities (lattice constant $a$) plotted against the equivalent spherical radius of the void. The light green dashed lines are calculated using the Tao-Eldrup model with different electron layer widths $\Delta$R. The dark blue line is the Tao-Eldrup curve for $\Delta$R = 3.14~a.u. (1.66 $\AA$) commonly used to fit experimental results. The red line is the Extended Tao-Eldrup model including excited Ps states according to a Boltzmann distribution at room temperature~\cite{CP_Goworek}. The horizontal dotted line marks the self-annihilation rate of o-Ps.} 
\label{fig8}
\end{center}
\end{figure}

The Ps-matter interaction energy, shown in Fig.~\ref{fig9}, is larger for denser crystals. It decreases toward larger voids, but remains positive because of the weakness of the van der Waals attraction. For voids larger than \rsph{} $>$ 6~a.u. the interaction energies fall into a single curve, irrespectively of the lattice parameter. 

Fig.~\ref{fig8} shows the pick-off annihilation rates \po{} against the free volumes of voids inside solid He of different densities. Not surprisingly, \po{} in a solid with so low polarizability as He is lower than $\lambda^{po}_{TE}$ parametrized for molecular materials and it is even below the self-annihilation rate of o-Ps for \rsph{} $\geq$ 8~a.u. They lie on a single curve independently of the lattice parameter. The pick-off annihilation rates for voids with equivalent spherical radii larger than 6~a.u. can be well approximated by the TE model with $\Delta$R=0.7~a.u. For small voids $\Delta$R is slightly larger because the positron is less localised inside the void and it annihilates with a larger probability in the bulk. $\Delta R$ is smaller in solid He than in most atomic molecular materials where the van der Waals attraction on Ps is stronger and can counteract the repulsion exerted by the core. 

Rytsölä et al.~\cite{JPB_Rytsola} measured o-Ps pick-off annihilation rates between 0.012~1/ns and 0.018~1/ns for solid He. According to Fig.~\ref{fig8} these values correspond to $r_{sph}$ = 6-7~a.u. Rytsölä et al. estimated the void radii for liquid $^4$He at low temperatures from the measured pick-off annihilation rates by assuming that Ps gets trapped in a spherical well of finite depth and obtained values between 6~a.u. and 9~a.u. The same approach yields for organic liquids values ranging between 7~a.u. and 11~a.u.~\cite{Book_Mogensen}. The \rsph{} has been estimated for a wide range of polymers using the TE model and $\mathrm{\Delta R}$=1.66~\AA (3.14~a.u.) giving values between 3.5~a.u. and 8~a.u. 
\begin{figure*}
\begin{center}
\includegraphics[width=16cm]{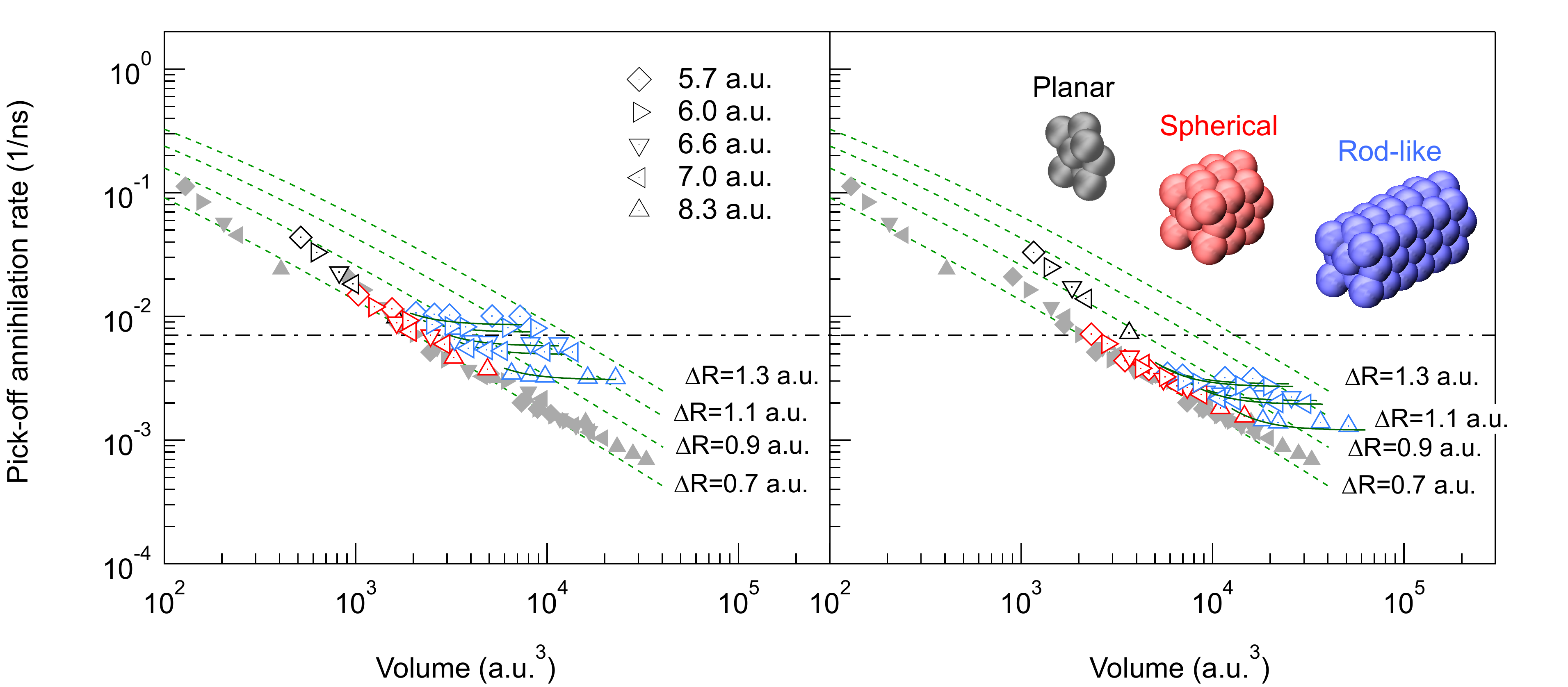}
\caption{(Color online) Ps pick-off annihilation rates for rod-like voids in solid He of different densities (lattice constant $a$). The rod axes are along the [2-1-10] direction and the results for cross-sections of 4 (left panel) and 9 (right panel) missing atoms are plotted versus the open volume. The colour of the marker describes the morphology of the void: planar (black), quasi-spherical (red) or rod-like (blue). Full grey markers are the annihilation rates in spherical voids given in Fig.~\ref{fig8}. The lines show the annihilation rates for spherical (green dashed lines) and rectangular (green full lines) voids within the TE model. Finally, the horizontal dotted-dashed line is the self-annihilation rate of o-Ps.} 
\label{fig13}
\end{center}
\end{figure*}

\subsection{Non-spherical voids}
We have also investigated how the morphology of the voids affects the Ps pick-off annihilation rates by introducing rod-like and planar voids in the supercells with lattice parameters ranging between 5.7~a.u. and 8.3~a.u. We want to study general effects of the void geometry, motivated by the fact that open volume pockets in porous materials, polymers and biological matter are rather long or flat cavelike than spherical. 

The rod axes are aligned along the [2-1-10] direction and they have rectangular cross-sections formed by 4 and 9 missing atoms. We have removed from 1 to 6 and 10 atom planes and finally we have also modeled an infinitely long rod-like void. The annihilation rates at the voids are plotted against their free volume in Fig.~\ref{fig13}. The free volumes of the infinitely long voids correspond to the volume within the supercell. 

The pick-off annihilation rates when a single atom plane is missing, i.e. for planar voids, are not sensitive to the size of the cross-section. They are, instead, given by the shortest dimension across the plane. When the length of the rod is comparable to the width of the cross section, the annihilation rates are comparable to the quasi-spherical voids of similar volume given in the previous section. Finally, for elongated rod like voids the annihilation rate saturates to a value depending only on the lattice parameter irrespectively of the free volume. The TE model for rectangular voids with the same cross-section of the atomistic calculation and $\mathrm{\Delta R}$=0.7 (full green lines in Fig.~\ref{fig13}) describes properly the annihilation rates in elongated voids. 
\begin{figure*}
\begin{center}
\includegraphics[width=16cm]{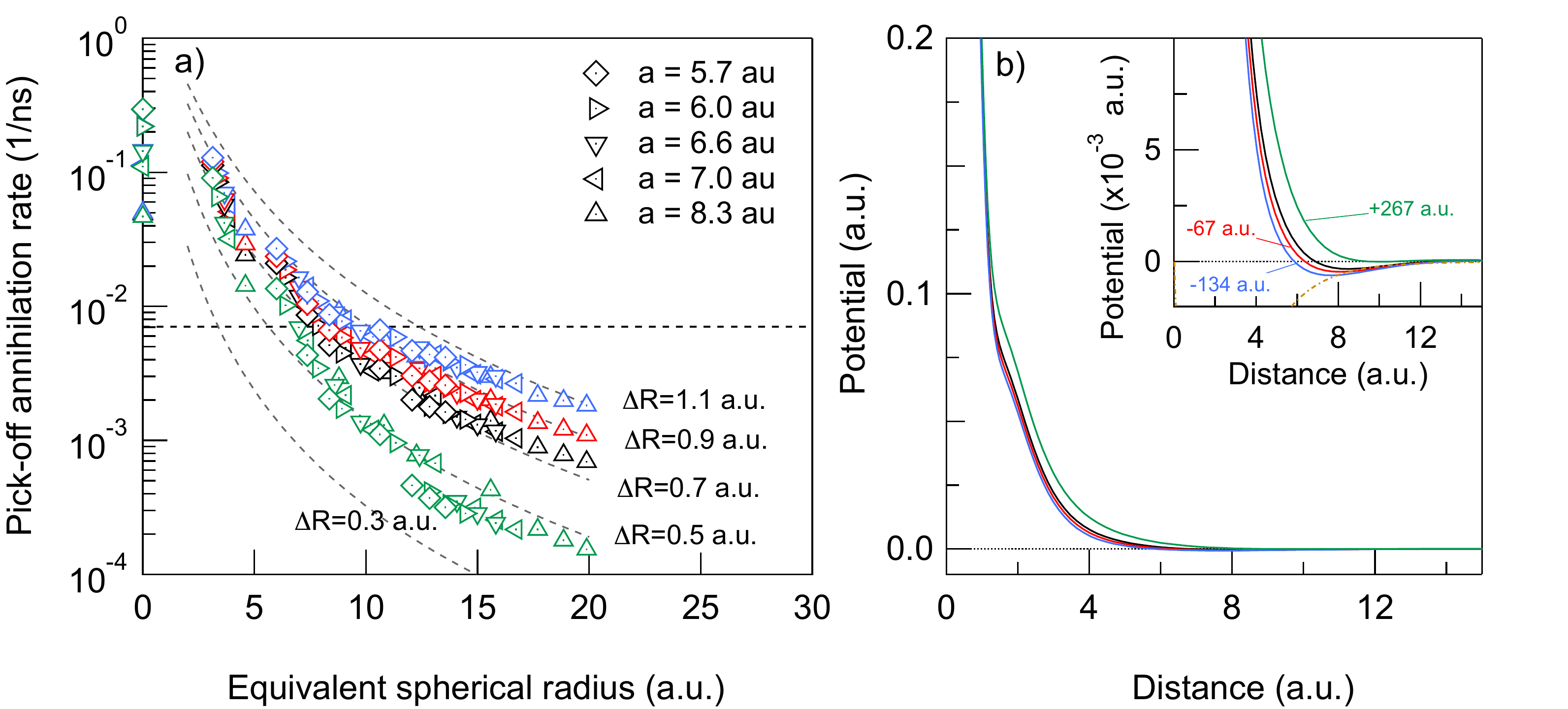}
\caption{(Color online) Modified van der Waals interaction. a) Ps pick-off annihilation rates at the bulk and inside the voids are plotted versus the equivalent spherical radius of the void. The original \Veff{} (black markers) and the modified \Veff{} with $\Delta$C$_6$ = -67 (red markers), $\Delta$C$_6$ = -134 (blue markers) and $\Delta$C$_6$ = +267.4 (green markers) are shown. The dashed grey lines are obtained from the TE model for spherical voids. The dashed horizontal line marks the self-annihilation rate of o-Ps. b) The original \Veff{} and the potentials with the modified van der Waals term. The inset shows a detail of the potential well. The dash-dotted orange line fits the van der Waals attractive contribution of \Veff{} as explained in the text.} 
\label{fig15}
\end{center}
\end{figure*}

\subsection{Effect of the dispersion interaction}
Although the attractive dispersion interaction felt by Ps is relatively weak in comparison with the Ps kinetic energy, it can be important on the low pick-off annihilation rate of Ps inside large voids. We have studied the effect by adding to \Veff{} an extra van der Waals interaction term V(R) =  g($\alpha$R) $\Delta$C$_6$/R$^6$ where R is the distance to the nucleus and g($\alpha$R) is a damping function~\cite{JCP_Tang} $g(\alpha R) = 1-\{\sum_{k=0}^6(\alpha R)^k/k!\}\exp(-\alpha R)$ which depends on the exponential decay parameter $\alpha$ of the low energy part of the \Veff{} potential. We obtain $\alpha$=0.9~a.u.$^{-1}$ by fitting \Veff{} between R = 2~a.u. and 4.5~a.u. to an exponential function. The van der Waals contribution in \Veff{} can be fitted with C$^0_6$ = -270. The resulting van der Waals term is plotted in the inset of Fig.~\ref{fig15} b). This value is roughly 20 times larger in magnitude than the value obtained by Mitroy and Bromley~\cite{PRA_Mitroy4}, -13.34, for the van der Waals coefficient for Ps interacting with He. However, the van der Waals coefficient for the interaction of the Ps atom as considered by Mitroy and Bromley cannot be directly compared to our potential for the positron within Ps. 

We have calculated the pick-off annihilation rates in spherical voids adding to \Veff{} a dispersion potential term with $\Delta$C$_6$ = -67 and $\Delta$C$_6$ = -134. Finally, we subtracted the dispersion term of \Veff{} by setting $\Delta$C$_6$= 270. The original \Veff{} and the modified potentials are plotted in the right panel of Fig.~\ref{fig15} and the corresponding annihilation rates in the left panel. Overall, the strengthened (weakened) van der Waals interaction has a negligible effect in the bulk but it increases (decreases) the pick-off annihilation rates of Ps inside the voids. 

The TE model fails to describe the attraction of the dispersion near the void wall. The attraction exerted by the wall is important for voids of \rsph{} $>$ 10~a.u. where the annihilation rate is higher than the TE model predicts. For the strongest dispersion that we considered the electron-layer thickness $\Delta R$ of the TE model that best fits the data ranges between 0.7~a.u. for voids of \rsph{} $\sim$ 5~a.u. and 1.1~a.u. when \rsph{} $\sim$ 20~a.u. On the other hand, when the dispersion attraction is subtracted from \Veff{} the annihilation rates are well described by the TE model for voids larger than 8~a.u. with $\Delta$R = 0.4-0.5~a.u. These trends are in agreement with the fact that the semiempirical $\Delta$R values for plastic materials and zeolites with stronger dispersion forces are much larger, of the order of 3-3.6 a.u. 

Our calculations show that the annihilation rates will depart from the prediction of the TE model for large voids. The void sizes estimated by the TE model using the measured Ps pick-off annihilation rates will be biased toward small values. As an example, when the annihilation rate is 0.0018~1/ns, according to the calculation with the strongest dispersion interaction the void radius are \rsph{} $\sim$ 20~a.u. but the TE model with $\Delta R$ = 0.7~a.u. predicts \rsph{} $\sim$ 13~a.u. In molecular materials (polymers and liquids) the sizes predicted for large voids can be severely underestimated due to their stronger van der Waals Ps-molecule interaction. More detailed calculations are needed for molecular systems to address this important problem in detail. 

\section{Conclusions}~\label{sec5}
We have used a full correlation single particle potential for the description of Ps states in matter. As a starting point, we have studied the HePs system and obtained the interaction potential \Veff{} by inverting a single particle Schr\"odinger equation. 
\Veff{} is characterised by a repulsive long-range exponential tail ($\alpha$ = 0.9~a.u.$^{-1}$) induced by the electron-electron exchange repulsion of the closed shell atom and the high zero-point energy of the light Ps atom. The agreement of the single particle potential elastic scattering parameters with the many-body values for momenta $\le$0.1~1/a.u. show that \Veff{} also describes the low energy correlations of quasi-thermalized Ps. The electron-electron and the positron-nucleus Coulomb repulsion dominates the repulsive core of \Veff{}. While positive ions exert a strong repulsion on the positron, the weaker repulsion of negative ions allows for an attractive potential well in unbound states suggesting that negative partial charges of polar molecules can attract Ps and enhance the pick-off annihilation rate. 

We introduce a model for Ps in hcp solid $^4$He using the superposition of atomic \Veff{} for the total  potential and use it to calculate the distributions and pick-off annihilation rates in voids of different geometries and sizes. We use the model to compare the annihilation rates calculated for spherical voids to the annihilation rates measured by Rytsola et al.~\cite{JPB_Rytsola}. for $^4$He at extremely low temperatures ($<$ 20~K) and conclude that HePs repulsion results in bubbles with radii of 6-7~a.u. This result is in good agreement with the estimation from a simple square-well model used in Ref.~\onlinecite{JPB_Rytsola}. 

Our results for spherical voids with 6 a.u. $<$ $r_{sph}$ $<$ 15 a.u. can be approximated by the TE model used widely to analyse positron annihilation data for polymers, liquids, and zeolites. The pick-off annihilation rates of planar and rod-like voids are also well described by the TE model. The volumes predicted by the spherical TE model for small (\rsph{} $<$ 6~a.u.) and large (\rsph{} $>$ 15~a.u.) open-volume pockets are underestimated because the TE model omits the annihilation beyond the surface layer of the void and the Ps-He dispersion attraction, respectively. This conclusion can be generalized to other materials in which Ps is formed. 

The pair-wise \Veff{} single-particle potentials can be used to study the properties of Ps in condensed matter without using computationally expensive many-body techniques. Generalized to other atoms than He and molecules such as methane and water, it will constitute a big advancement in the analysis of the lifetime experiments of o-Ps in molecular materials. 

\begin{acknowledgments}
This work was supported by the Academy of Finland through individual fellowships and Centres of  Excellence Program (project number 251748). We acknowledge the computational resources provided by the Aalto Science-IT project. Thanks are due to K. Varga for providing us the ECG-SVM code used in this work, to A.~P. Seitsonen for insightful discussions about the technical aspects of the calculations, and to K. Rytsölä for valuable discussions about the experimental works. 
\end{acknowledgments}


%

\end{document}